\documentclass[10pt]{bmc_article}

\usepackage{cite} % Make references as [1-4], not [1,2,3,4]
\usepackage{url}  % Formatting web addresses  
\usepackage{ifthen}  % Conditional 
\usepackage{multicol}   %Columns
\usepackage[utf8]{inputenc} %unicode support
\usepackage{float}
\usepackage{graphicx}
\usepackage{amssymb}
\usepackage{algorithmic}
\usepackage{algorithm}
\usepackage[english]{babel}
\usepackage{blindtext}
\newboolean{publ}

\newenvironment{bmcformat}{\baselineskip20pt\sloppy\setboolean{publ}{false}}{\baselineskip20pt\sloppy}

\begin{document}
\begin{bmcformat}


\title{Gene set bagging for estimating replicability of gene set analyses}
 
%%%%%%%%%%%%%%%%%%%%%%%%%%%%%%%%%%%%%%%%%%%%%%
%%                                          %%
%% Enter the authors here                   %%
%%                                          %%
%% Ensure \and is entered between all but   %%
%% the last two authors. This will be       %%
%% replaced by a comma in the final article %%
%%                                          %%
%% Ensure there are no trailing spaces at   %% 
%% the ends of the lines                    %%     	
%%                                          %%
%%%%%%%%%%%%%%%%%%%%%%%%%%%%%%%%%%%%%%%%%%%%%%

\author{Andrew E. Jaffe$^{1,2}$ \and
	John D. Storey$^3$ \and
	Hongkai Ji$^1$ and
	Jeffrey T. Leek\correspondingauthor$^1$
         \email{Jeffrey T Leek\correspondingauthor: jleek@jhsph.edu}
 }

%%%%%%%%%%%%%%%%%%%%%%%%%%%%%%%%%%%%%%%%%%%%%%
%%                                          %%
%% Enter the authors' addresses here        %%
%%                                          %%
%%%%%%%%%%%%%%%%%%%%%%%%%%%%%%%%%%%%%%%%%%%%%%

\address{%
    \iid(1)Department of Biostatistics, Johns Hopkins Bloomberg School of Public Health, Baltimore MD 21205 \\
    \iid(2)Lieber Institute for Brain Development, Johns Hopkins Medical Campus, Baltimore MD 21205 \\
    \iid(3)Lewis-Sigler Institute and Department of Molecular Biology, Princeton University, Princeton, NJ 08544
}%

\maketitle

%%%%%%%%%%%%%%%%%%%%%%%%%%%%%%%%%%%%%%%%%%%%%%
%%                                          %%
%% The Abstract begins here                 %%
%%                                          %%  
%% Please refer to the Instructions for     %%
%% authors on http://www.biomedcentral.com  %%
%% and include the section headings         %%
%% accordingly for your article type.       %%   
%%                                          %%
%%%%%%%%%%%%%%%%%%%%%%%%%%%%%%%%%%%%%%%%%%%%%%

\begin{abstract}
\section*{Background}
Significance analysis plays a major role in identifying and ranking genes, transcription factor binding sites, DNA methylation regions, and other high-throughput features for association with disease. We propose a new approach, called \emph{gene set bagging}, for measuring the stability of ranking procedures using predefined gene sets. Gene set bagging involves resampling the original high-throughput data, performing gene-set analysis on the resampled data, and confirming that biological categories replicate. This procedure can be thought of as bootstrapping gene-set analysis and can be used to determine which are the most reproducible gene sets. 
\section*{Results} Here we apply this approach to two common genomics applications: gene expression and DNA methylation. Even with state-of-the-art statistical ranking procedures, significant categories in a gene set enrichment analysis may be unstable when subjected to resampling. 
\section*{Conclusions}
We demonstrate that gene lists are not necessarily stable, and therefore additional steps like gene set bagging can improve biological inference of gene set analysis. 
\section*{Key Words}
Gene Set Analysis, Gene Expression, DNA Methylation, Gene Ontology      
\end{abstract}

\ifthenelse{\boolean{publ}}{\begin{multicols}{2}}{}

%%%%%%%%%%%%%%%%
%% Background %%
%%
\section*{Background}

The biology of many organisms is organized naturally as a series of diverse pathways, and the genetic landscape of cells is no different - genes also group together in pathways to perform specific functions \cite{hood2004}. Human health depends on the functionality of these pathways; de-regulation at the pathway level may be more important for diseases like cancer than de-regulation of specific genes \cite{vogelstein2004}. The most common statistical approach for identifying pathways of interest in a high-throughput experiment is to perform a significance analysis gene-by-gene and then summarize the significant hits using gene set or gene pathway analyses. Each pathway or gene-set analysis is performed once on the entire data set. However, there is variability in the identified gene sets due to both the instability in gene rankings from the original gene ranking analysis and from the pathway/set analysis. 

Here we propose a new approach to evaluate the stability of biological inference drawn from an experiment. Our approach, called gene set bagging, performs a resampling of the entire discovery algorithm - significance analysis and gene set enrichment - to identify the most stable and reproducible enriched gene sets. We perform resampling by drawing an equal number of samples with replacement from the full (observed) dataset, performing a significance analysis followed by gene set analysis, and then identifying which sets are enriched. We can identify which observed gene sets are consistently enriched in resampled data, and compute the gene set replication probability ($R$), a measure of gene set stability based directly on the biological quantity of interest, representing the probability that an observed gene set will be enriched in future experiments. 

The replication proportion ($R$) has some important advantages over the traditionally-reported p-value for summarizing gene set enrichment. The structure of the gene set testing problem is fundamentally different than other multiple hypothesis testing problems - correlations between genes, different gene set sizes, and different levels and fraction of differential expression within gene sets makes the hypotheses fundamentally not comparable with standard significance testing \cite{efron2007testing,gatti2010}. We therefore instead propose to estimate directly the probability that a gene set will replicate, as in this more complicated multiple testing scenario, an estimate of the probability of replication may be of more interest than a measure of statistical significance. Lastly, given the emphasis for replication in genetics/genomics studies, this replication proportion may be another metric for directing molecular validation of important biological processes involved in human disease. 

We perform our gene set bagging method on two genomics measurements: gene expression and DNA methylation. Even after adjusting the genomic data for potential batch effects, we demonstrate that some significant gene sets fail to replicate well, yet other non-significant sets have high replication rates. The results for these different genomic technologies suggest that the signal and noise structure of the specific genomic data type contribute greatly to stability of gene sets. We use a simulation study to assess replication across two simulated datasets, and evaluate the concordance between replication probability ($R$) and the traditionally-reported significance metric (p-value). We show that the replication probability better quantifies the chance that a significant gene set will be consistent across studies, and the result of our analyses suggests that: (1) gene set enrichment analyses from a typical high-throughput study may be highly unstable, and (2) gene set bagging is a resampling approach for measuring the stability of gene sets and ensuring reproducibility of biological conclusions.

\section*{Methods}

Bagging, also known as bootstrap aggregating, is traditionally used for assessing the predictive accuracy and stability of prediction models \cite{hastieElements}. While bagging procedures have been used for differential expression analyses \cite{dudoit2002comparison}, here we introduce a new bagging procedure for significance analysis of gene sets called gene set bagging. This procedure can be useful for both evaluating significance rankings and also for describing the most reproducible genes and biological gene sets within genomics experiments in a platform-independent fashion. 

For gene set $\l$, the goal is to estimate: $$R_l = {\rm Pr}({\rm Gene \; set \; l \; will \; be\; significant \; in \; a \; new \; study}).$$
The quantity $R_l$ is useful as a measure of the stability of the significance of an identified gene set. Gene sets are frequently used to interpret the biological results of studies, so it is important to know if the biological "story" would change if the study was repeated. This is particularly true since gene set analysis is subject to errors in annotation, variation due to technological noise, and variation due to biological noise. 

As an example of our general approach, we focus on a cigarette smoking dataset (further explained in the following Datasets and Implementation section), which examined gene expression differences associated with smoking exposure in 40 smokers and 39 never-smokers. We define gene expression measurements $m_{ij}$ for each of $j=1,\ldots,79$ samples over $i=1,\ldots,M$ genes/probes (corresponding to gene $g_i$) and a covariate of interest per sample ($z_j \in [current smoker, never smoker]$). We first want to identify differentially expressed genes between the two outcome groups, so we calculate an empirical Bayes t-statistic and resulting p-value for each gene \cite{smyth2004limma}. We can call any gene significant if $\alpha < 0.05$ (or, alternatively, we can assume $\alpha$ to be the family-wise error rate as to control for multiple testing), and look for enrichment in $L$ predefined gene sets. Each gene set gets a p-value ($p_l$), reflecting the degree of enrichment. The prevalent wilcoxon mean rank gene set enrichment test \cite{michaud2008} available in the limma Bioconductor package \cite{limma} and traditional hypergeometric test were used as enrichment tests for each dataset. 

We can then perform gene set bagging using $B=100$ iterations. In each iteration ($b$), we resample the 40 smokers and 39 smokers separately with replacement. Each gene or probe gets a p-value via calculating a t-statistic in the resampled data, and these statistics are passed to gene set analysis algorithms to produce a p-value of enrichment for each gene set ($p_l^{b}$), which are stored in the first column of a [\#gene sets by $B$] matrix.

In the next iteration, we resample again and create a different resampled dataset, and get another set of p-values that are put in the second column of the storage matrix. We fill in the columns of the matrix if we do this 100 times. For each row, which represents a single gene or probe, we count the number of times each subsampled p-value ($p_l^{b}$) is less than $\alpha$ (here, 0.05), and divide it by the number of iterations (B), resulting in an estimate of the replication probability for that gene set ($\hat{R}_l$).

Regardless of the application, estimated replication probabilities ($\hat{R}$) are between 0 and 1, where 0 means that the gene set always had a p-value greater than $\alpha$ in every iteration, and 1 means that the category always had a p-value less than $\alpha$ in each iteration. For analyses where the gene ranking is stable and the gene set calculation is stable, the replication probability will be higher. This estimate of replication assesses the stability of the gene sets, and might be a better estimate of biological reproducibility than the traditionally reported p-values. Our goal is to identify the more stable set of gene sets, akin to Meinshausen, and B\"{u}hlmann 2010 \cite{meinshausen2010} in selecting a more stable set of covariates in a regression model.

%% for set

\floatstyle{ruled}
\newfloat{algorithm}{htbp}{loa}
\floatname{algorithm}{Algorithm}
\begin{algorithm}
\begin{algorithmic}
\STATE 1. Estimate a test statistic for each gene $\hat{T}_i$\\
\STATE 2. Use the test statistics to calculate a P-value for each gene set, $p_l, \; l = 1 \ldots, L$, using any standard gene set analysis algorithm. \\
\STATE 3. For ($b \in 1,\ldots,B$):\\
\hspace{.25 cm} i. Resample individuals within outcome groups   \\
\hspace{.25 cm} ii. Estimate a bootstrap test statistic for each gene $\hat{T}^{*b}_i$\\ 
\hspace{.25 cm} iii. Use the test statistics to calculate a bootstrap p-value \\
\hspace{.25 cm} for each gene set, $p^{*b}_l, \; l = 1 \ldots, L$, using any standard gene\\
\hspace{.25 cm} set analysis algorithm.
\STATE 4. Estimate the replication probability $\hat{R}_l = \frac{\sum_{b=1}^B \mathbb{I}[p_l^{*b} < \alpha]}{B}$ for each gene set. 
\end{algorithmic}
\caption{Gene set bagging procedure}
\label{set_alg}
\end{algorithm}

\subsection*{Datasets and Implementation}

\subsubsection*{Simulated Data}

We designed two simulation studies to assess different properties of the replication probability based on the Affymetrix Human Genome 133 Plus 2.0 gene expression microarray. Basing the simulation on an existing array design, with probes annotated to genes that were already mapped to gene ontology categories, allowed us to realistically add differential expression signal to specific gene sets. We first selected a random sample of 100 gene sets to use in our simulation, which corresponded to 2288 unique genes. Then, for each simulation, we selected first 100 and then 500 genes (generally denoted $G$) to insert signal into, via the following model:

$$ m_{ij} = \beta_0 + \beta_{i} z_j + \epsilon_{ij} $$

where $\epsilon_{ij} \sim N(6,1)$ and  $\beta_i \sim N(1,0.5)$ if $g_i \in G$, and $m_{ij}$ and $z_j$ (defined above) correspond to the expression value and binary outcome respectively. 

In the first simulation, we generated 1000 datasets, where each consisted of 100 individuals (50 cases and 50 controls). For each dataset, we inserted signal into 100 genes and computed the observed p-value ($p_l$) and then the replication probabilities ($R_l$) for each gene set $l$. In the second simulation we generated 100 pairs of datasets, where each dataset contained 50 individuals (25 cases and 25 controls), and inserted signal into 500 genes, and computed observed p-values and replications probabilities for each gene set. 

\subsubsection*{Gene expression: cigarette smoking data}

We tested the gene set bagging method in a differential expression analysis with data obtained from Gene Expression Omnibus (GSE17913). This study examined the effect of cigarette smoking on the oral epithelial transcriptome by comparing buccal biopsies in 39 never-smokers with 40 active-smokers using the Affymetrix Human Genome U133 Plus 2.0 microarray \cite{boyle2010}. We processed the raw CEL files using the RMA algorithm to perform intra-array normalization and then performed quantile normalization to adjust for between-array biases \cite{irizarry2003}.

We performed surrogate variable analysis (SVA) to adjust for potential batch effects \cite{leek2007,leek2008general}. Briefly, this approach identifies the number of right singular vectors that are associated with more variation than expected by chance, and then in the subsets of genes driving this variation, constructs a 'surrogate' variable for each subset. These surrogate variables are then included as covariates in our differential expression analysis (so that the model becomes: $m_{ij}=\beta_i z+ \beta_{SV} SV+ \epsilon_{ij}$). 

We identified differentially expressed genes comparing cases and controls while controlling for the surrogate variables using an empirical Bayes approach \cite{smyth2004}. To determine statistical significance, resulting p-values were converted to q-values to control for the false discovery rate \cite{storey2003} and all transcripts with q-values less than 0.05 were considered significant. We performed a full gene ontology analysis, and then ran the gene set bagging algorithm.

\subsubsection*{DNA methylation: brain tissue}

We believe this approach to be generalizable to most genomics platforms, and first tested this hypothesis using DNA methylation data processed on the Illumina HumanMethylation27 platform using freely available data from GEO \cite{gibbs2010} [GSE15745]. Our analysis utilized DNA methylation data from a recent paper that assessed quantitative trait loci using methylation and expression data in four different brain tissues. Previous work has identified that DNA methylation signatures can distinguish brain tissues, and might play a role in determining and stabilizing normal brain differentiation \cite{ladd2007}. We conducted our gene set bagging algorithm on the differential DNA methylation analysis between the frontal and temporal cortices. Detailed preprocessing information can be found in the supplementary material.

We performed the full differential methylation analysis comparing 131 front cortex and 126 temporal cortex samples, adjusting for plate number, tissue bank site, sex, and age, using the exact same approach as the gene expression example (with and without SVA, then empirical Bayes and multiple testing correction on each). All probes with q-values less than 0.05 were considered significant. We performed a full gene ontology analysis on the gene associated with each probe (from the annotation table), and ran the gene set bagging algorithm. 

\section*{Interpretation}

\subsection*{The replication probability ($R$) reflects stability}

The interpretation of the replication probability reflects the underlying stability of each outcome group. Suppose we observe a p-value for a gene set that we call significant at significance level $\alpha$. The replication probability estimate is defined as the fraction of times that feature is significant at the same $\alpha$ level in resamplings of the original data. If we called all p-values significant at $\alpha=0.05$, $R=0.8$ means that feature had a p-value less than 0.05 in 80\% of resampling datasets. If the statistical signal is stable, significant features will have high replication probability estimates, and non-significant features will have low replication probability estimates ($R \approx 0$), because the resampled data should be representative of the overall population. 

To better understand how the replication probability therefore addresses stability, suppose we perform a gene expression experiment, and further study two gene sets: Set A with $p=0.001$ and Set B with $p=0.2$. We then calculate the replication probability for these two gene sets, and want to interpret the results. Consistency between the replication probability and p-value means that the direction of statistical inference is identical: high replication probabilities with low p-values are consistent. 

% jaffe delete all the crazy comma induced problems. why you copying from word anyway, son?
\begin{table}[htbp]
  \centering
  \caption{Interpretations of sequential replication probabilities ($\hat{R}$) for two different experiment features.}
    \begin{tabular}{|c|r|}
	\hline
    \textit{\textbf{R}} & \textbf{"Feature A (p = 0.001) is significant,} \\
	\hline
    0     & but very inconsistent” \\
    0.25  &  but very inconsistent" \\
    0.5   & inconsistent" \\
    0.75  & somewhat consistent" \\
    1     & very consistent" \\ 
\hline
\\
\hline
     \textit{\textbf{R}}      & \textbf{"Feature B (p = 0.2) is non-significant,} \\
    \hline
	0     & very consistent" \\
    0.25  & somewhat consistent" \\
    0.5   & inconsistent" \\
    0.75  & very inconsistent" \\
    1     & very inconsistent" \\
\hline
    \end{tabular}%
  \label{tab1}%
\end{table}%

To understand the apparently counterintuitive interpretations, we will focus on the row in Table \ref{tab1} where $R=1.0$. Feature A is extremely stable because the significance of this feature seems constant throughout resampling. Feature B is also extremely stable because the resampling inferences are relatively non-variable, but the replication probability is very inconsistent with the p-value. Interpreting stability and consistency through the replication probability therefore requires observed p-values for gene sets. However, it is important to remember that the replication probability is a function of the statistical significance level ($\alpha$); algebra can demonstrate that $R$ increases as $\alpha$ increases. 

\subsection*{$R$ estimates the probability a gene set will be significant in a repeated study}

We simulated 1,000 identical data sets (as described in the Datasets section; Simulation 1), these data sets represent repeated experiments performed under the same conditions. We spiked in specific genes as differentially expressed in these simulated data sets. We then performed gene set analysis using both the hypergeometric and Wilcoxon tests and calculated the replication probability estimates for each gene set in each of the 1,000 simulated studies. The average replication probability estimate across all 1,000 repeated studies very closely approximates the frequency that a gene set is observed to be significant in those 1,000 studies (Figures \ref{fig1}A and \ref{fig1}B). In other words, the estimate of the replication probability is close to the probability a gene set will be significant in a future study.

\subsection*{Replication adds biological interpretability}

In the gene expression dataset (Figure \ref{fig2}), there were 8 GO categories with p $>$ 0.05 and R $>$ 0.8 under the hypergeometric test, including sets associated with phosphorylation (GO:0006468, GO:0016310), a process affected by cigarette smoking \cite{anto2002} and regulation of metabolic processes (GO:0019222, GO:0044267). Similarly, examining the categories associated with DNA methylation differences across brain tissue that had at least moderate replication and non-significant p-values demonstrates support for the gene set bagging approach as well as the shortcomings of relying on strict p-values cutoffs for gene ontology analysis (Figure \ref{fig3}). Several biologically plausible GO categories for a comparison of methylation differences in brain tissues fell into the "marginally significant" bin of observed p-values between 0.05 and 0.1 but had consistent replication.  

Similarly, there were many smaller gene sets that had statistically significant p-values (p $<$ 0.05) but never appeared in any of the resampled datasets ($R$=0) in both the gene expression ($N=32$) and DNA methylation datasets ($N=12$). These represent very unstable gene sets, and should be interpreted with caution. Categories like the first set ($p > 0.05, R > 0.8$) would have been ignored in a traditional gene set analysis given their statistical significance measure, but might be biologically important to the question of interest. Likewise, gene sets in the second category ($p<0.05, R=0$) are probably less biologically meaningful even though they are "statistically significant".

\section*{Results}

\subsection*{$R$ correlates better with replication in independent datasets}

Besides identifying which gene sets are the most stable, we can also assess how well the replication probability ($R$) reflects biological replication by spiking-in gene set enrichment  across two independent simulated datasets (described fully in the Datasets section). We performed traditional gene ontology analysis on both datasets, obtaining p-values for each gene set and study calculated from the hypergeometric distribution, and then performed our gene set bagging algorithm on these same two datasets. There was very strong Spearman correlation between datasets across 100 simulation runs when all gene sets were considered regardless of whether the replication proportion (median=0.854, IQR: 0.826-0.876) or p-value (median = 0.836, IQR: 0.809-0.869) was used (Figure \ref{fig1}C). However, when only gene sets where at least 1 of 2 datasets was significant at $p < 0.05$ per simulation run, the replication proportion had much stronger correlation (median = 0.755, IQR =0.678-0.817) than the p-value (median = 0.535, IQR: 0.387 - 0.648) (Figure \ref{fig1}D). These results suggest that globally, there might not be a large difference between the replication proportion and the p-value, but when there is any signal in a particular gene set, the replication proportion better captures independent replication of that set in future studies. We also performed the more robust Wilcoxon rank rest on these simulated paired datasets, which also had less correlation between the resulting p-values than the replication probability (Figure \ref{fig1}E). There were many fewer significant gene sets by this enrichment approach than the hypergeometric test, and it was rare that both independent datasets within a simulation were significant at $p < 0.05$. The p-values derived from the Wilcoxon method therefore appear more conservative than collapsing each gene set into a 2x2 table and performing the hypergeometric test.

\subsection*{Relationship to the problem of regions}

The set of test statistics corresponding to genes within an individual set can be viewed as a multivariate random variable. When viewed in this way, a gene set is significant if the vector of test statistics falls into a multi-dimensional region defined by the significance threshold. The replication probability is then a first approximation estimate of the posterior probability a gene set will be significant, assuming a non-informative prior distribution on the vector of test statistics. This problem has been considered in the case of multivariate normal data \cite{efron1998problem} and for estimating confidence in inferred phylogenies \cite{felsenstein1985confidence}. As has been previously pointed out, this posterior probability is a reasonable first approximation to the posterior probability in question, but should not be interpreted as a frequentist measure of statistical significance \cite{efron1998problem,efron1996bootstrap}. 

As an example of the relationship between the bootstrap and a posterior probability, suppose $z_1,\ldots, z_n \sim N(\mu, \sigma^2)$. A non-informative prior distribution for the parameters $(\mu, \sigma^2)$ is the Jeffrey's prior \cite{jeffreys1946invariant}. The Jeffrey's prior for $\mu$ is an improper uniform prior across the real line and the Jeffrey's prior for $\sigma^2 \propto \frac{1}{\sigma^2}$. Using these prior distributions, the posterior distribution for $\mu$ is $N(\bar{z},\tau^2)$ where $\tau \sim InverseWishart_{n-1}((n s^2)^{-1})$ and $s^2 = \frac{1}{n} \sum_{i=1}^n (z_i - \bar{z})^2$. In this case, since $\mu$ is one dimensional, the $InverseWishart$ distribution is equivalent to an $InveseGamma$ distribution. Drawing bootstrap samples from the $z_i$ and recalculating the mean approximates sampling from the posterior distribution of $\mu$ (see supplemental R code). It is important to note that the variance of the posterior for $\mu$ is inflated compared to $\sigma^2$ assuming a frequentist model \cite{efron1996bootstrap,efron1998problem}. Note that the p-values these bootstrap samples should not be interpreted as measures of statistical significance, because they are no longer distributed uniformly.

\section*{Discussion}

We have developed a resampling-based strategy for estimating the probability a gene set will replicate as statistically significant. This direct approach to estimating replicability may be more useful than statistical significance for investigators who aim to identify stable and reproducible biological stories for their results. By utilizing outcome-based resamplings of the observed data, the reproducibility of gene sets can be quantified, represented by the replication probability $R$ of each gene set category across all subsamples. This approach can offer an additional metric beyond the p-value for identifying which biological pathways to follow-up. We have successfully applied this method to gene expression and DNA methylation under two commonly-used enrichment metrics: the hypergeometric test and the wilcoxon rank test, and demonstrate that many seemingly statistically significant GO categories fail to replicate consistently. A strength of our approach is the generalizability of this algorithm to most other genomics applications, including following bias-correcting approaches like SVA during the analysis, to assess the stability of results lists.

Overall, between the two most commonly used gene set enrichment measures, the Wilcoxon rank test appears more stable than the hypergeometric p-value, using simulated and real data. There were fewer inconsistent gene sets (significant either by $R$ or the p-value, but not the other), and the relationship between the replication probability and p-value was more precisely defined (Figure \ref{fig1}B and \ref{fig2}B).

Gene sets with high replication probabilities and low p-values represent statistically significant, stable, and consistent sets that might best represent the underlying biology within the experiment.  Given that most genomics studies require some form of external replication and that $R$ appears more correlated with replication in second follow-up study than p-values alone, we might also suggest following-up gene sets that have high replication probabilities ($R$) even if the p-values are marginally, or even non-, significant. From a practical perspective, the gene set bagging algorithm has been turn into the R package "GeneSetBagging", available through GitHub (https://github.com/andrewejaffe/GeneSetBagging). While defining a recommended cutoff seems counterproductive, as in different applications, users may choose different cutoffs depending on their resources for replication and how willing they are to be correct, a cutoff value of $R$ above 0.5 means that gene set is more likely to replicate than not, and could be used as a lower bound for replicability. 

Typical genomics practice often involves drawing the majority of biological conclusions of an experiment from the results of a gene set analysis without assessing the stability of the results. We envision replication probabilities used in conjunction with standard measures of statistical significance, as the emphasis on replication in genetics/genomics makes the replication probability a useful quantity to estimate and use in conjunction with p-values. We have demonstrated that gene lists are not necessarily stable, and therefore additional steps like gene set bagging should be undertaken to maximize the biological inference of a given study. 

\bigskip

%%%%%%%%%%%%%%%%%%%%%%%%%%%%%%%%
\section*{Author's contributions}
JTL and JDS designed the study, AEJ performed the analyses, AEJ HJ JDS and JTL wrote the manuscript.

%%%%%%%%%%%%%%%%%%%%%%%%%%%
\section*{Acknowledgements}

Funding: National Institutes of Health [grant numbers P50HG003233, R01HG005220], and a Johns Hopkins School of Public Health Faculty Innovation Award to J. Leek
 
%%  Bmc_article.bst  will be used to                       %%
%%  create a .BBL file for submission, which includes      %%
%%  XML structured for BMC.                                %%
%%  After submission of the .TEX file,                     %%
%%  you will be prompted to submit your .BBL file.         %%

\newpage
{\ifthenelse{\boolean{publ}}{\footnotesize}{\small}
 \bibliographystyle{bmc_article}  % Style BST file
  \bibliography{bibliography} }     % Bibliography file (usually '*.bib' ) 

%%%%%%%%%%%

\newpage

\begin{figure}
\begin{center}
\includegraphics[height=5in]{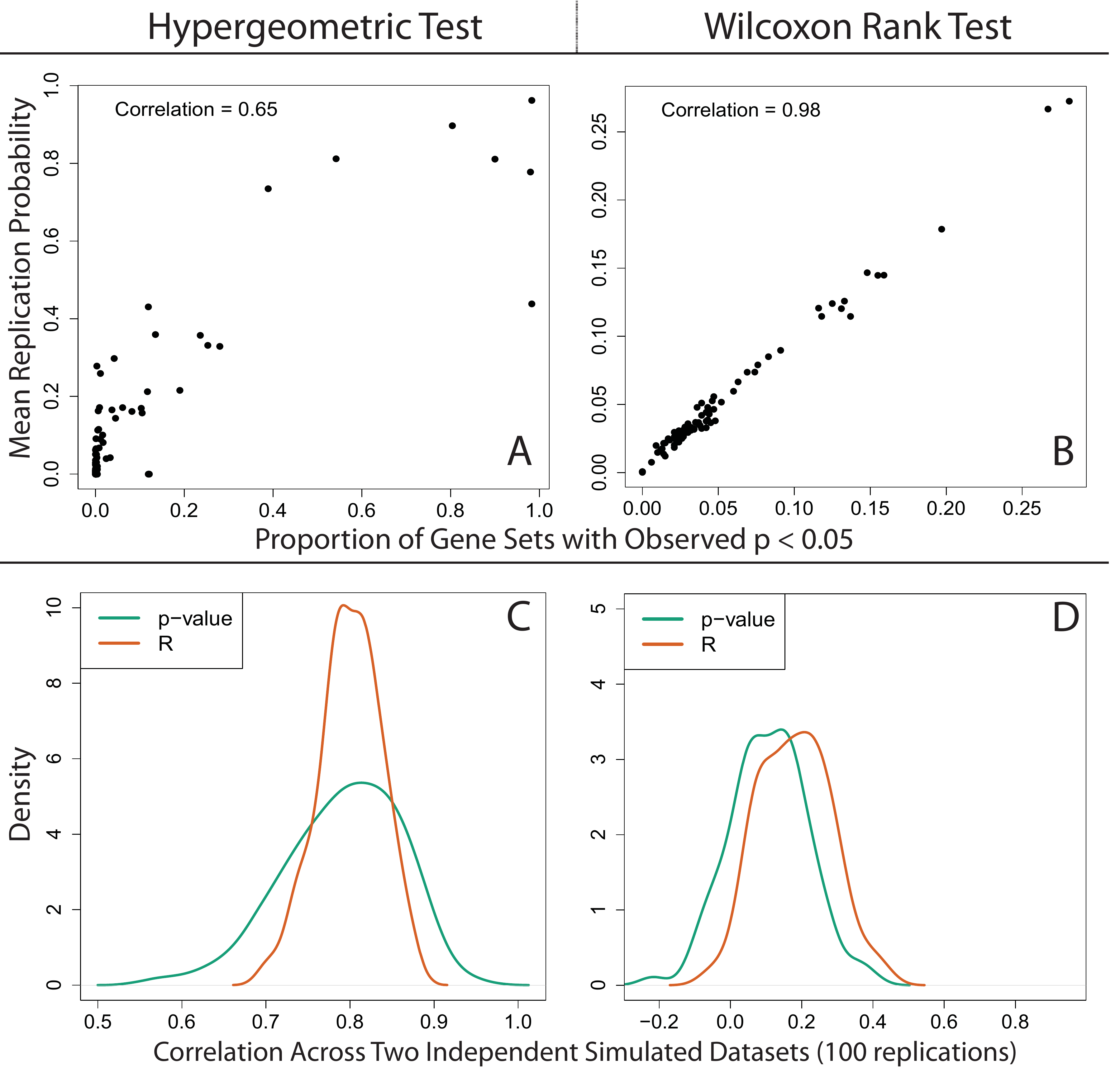}
\end{center}
\caption{Replicability assessed by simulation. Differential expression of specific genes was inserted into 1000 datasets, and both the observed p-values for the (A) hypergeometric and (B) Wilcoxon Rank tests and subsequent replication probabilities were calculated. The x-axis is the proportion of observed p-values that are less than 0.05 for each gene set  and the y-axis is the average replication probability for that gene set. Correlations were calculated using the Spearman method to avoid issues with non-linearity. Then, one hundred sets (2288 genes) were generated each with two independent simulated expression datasets, with differential expression signal inserted with some variability at 500 genes. Gene set tests were performed by the hypergeometric test, followed by gene set bagging, and distributions of Spearman correlations between independent datasets across 100 simulation runs for (C) all gene sets and (D) those significant in either independent dataset at $p < 0.05$. The replication proportion offers better correlation between independent datasets for significant gene sets, but similar correlation across all significant and non-significant gene sets, than the p-value for the hypergeometric test. Lastly, (E) overall correlation is improved for the replication proportion versus the p-value for the  Wilcoxon rank test when all gene sets are considered.}
\label{fig1}
\end{figure}

\newpage

\begin{figure}
\begin{center}
\includegraphics[height=3in]{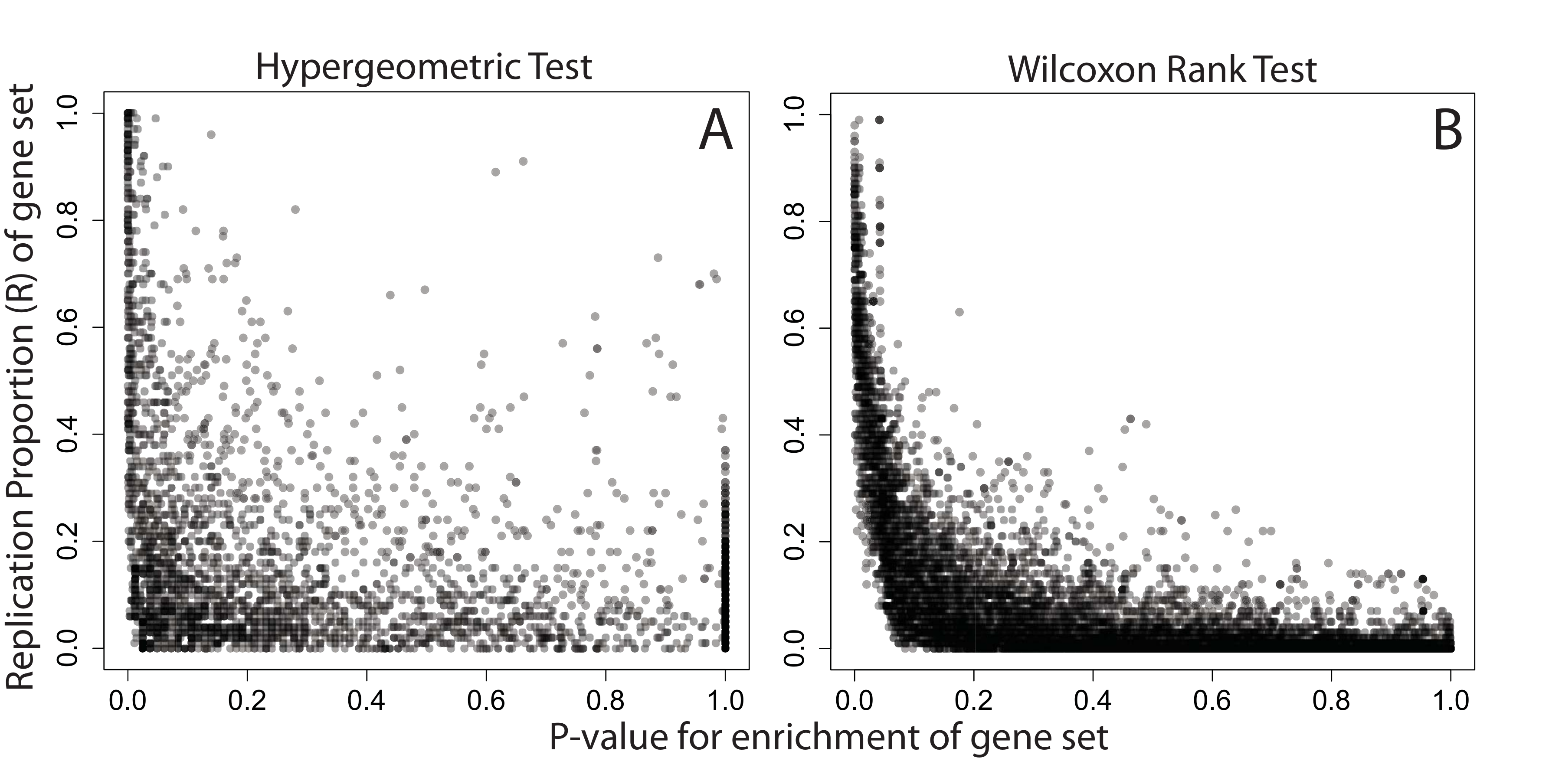}
\end{center}
\caption{Expression dataset gene set analysis. Gene set analyses were performed by the (A) hypergeometric and (B) Wilcoxon rank tests using gene sets defined by the Gene Ontology, and the replication of each gene set was assessed via our gene set bagging procedure (each point is one gene set). The relationship between the estimated replication probability ($\hat{R}$) and traditionally reported p-value appears much more stable using the Wilxocon rank test. }
\label{fig2}
\end{figure}

\newpage

\begin{figure}
\begin{center}
\includegraphics[height=3in]{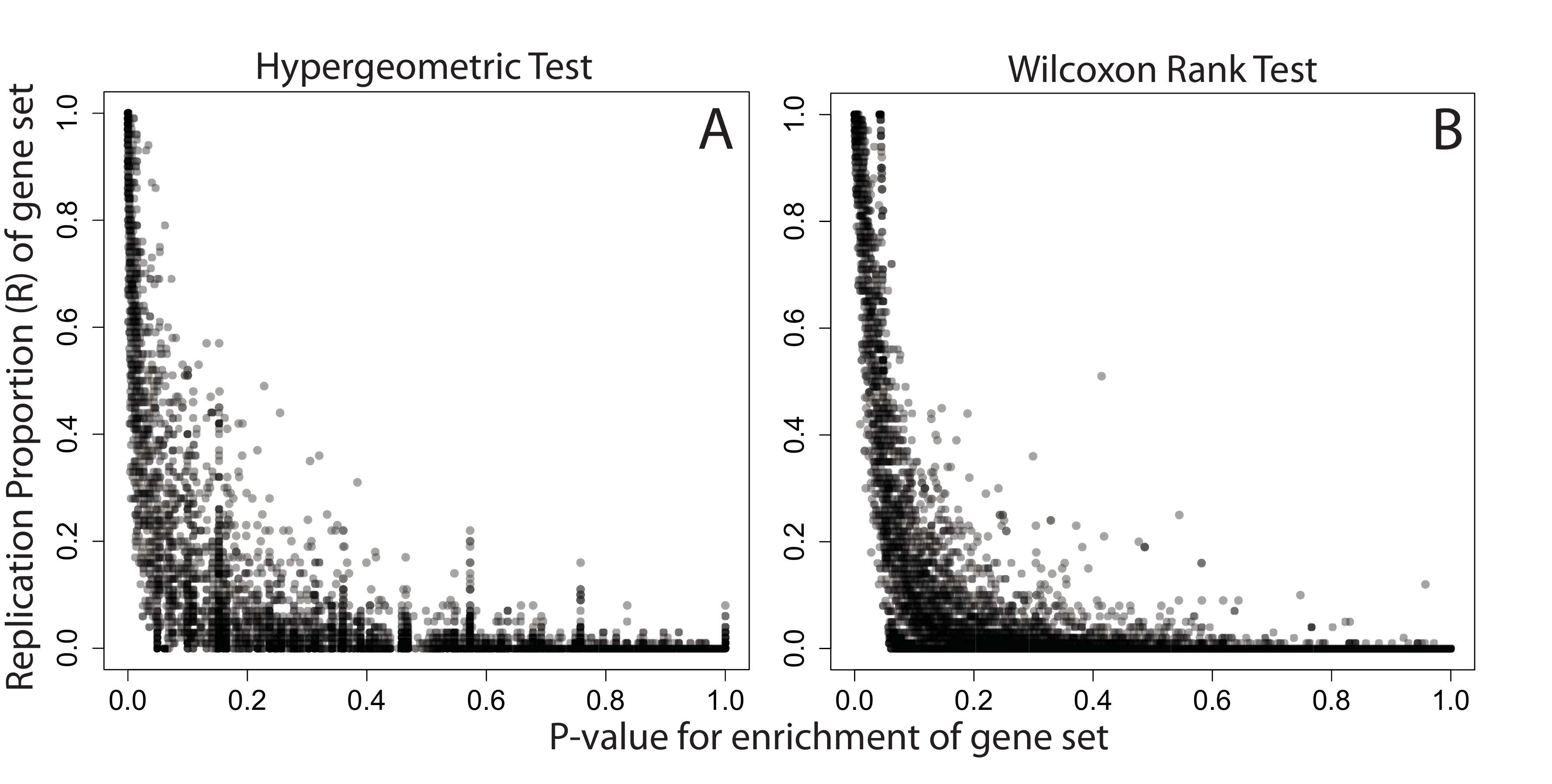}
\end{center}
\caption{DNA methylation dataset gene set analysis. Gene set analyses and gene set bagging were performed by the (A) hypergeometric and (B) Wilcoxon rank tests using gene sets defined by the Gene Ontology. The relationship between the estimated replication probability ($\hat{R}$) and traditionally reported p-value appears are now only slightly more stable by the Wilxocon rank test, which can be attributed to this dataset having approximately three times more samples under study. }
\label{fig3}
\end{figure}

\ifthenelse{\boolean{publ}}{\end{multicols}}{}

\end{bmcformat}
\end{document}